\documentclass{article}
\usepackage{stfloats,algorithmicx,multirow,latexsym}
\usepackage{spconf}
\usepackage{booktabs}
\usepackage{cite}
\usepackage[dvips]{graphicx}
\usepackage{amsmath}	
\usepackage{amssymb}    
\usepackage{amsfonts}   
\usepackage{mathrsfs}
\usepackage{array}
\usepackage{mdwmath}
\usepackage{mdwtab}
\usepackage{subfig}
\usepackage{fixltx2e}
\usepackage{soul}
\usepackage{verbatim}
\usepackage{url}
\usepackage{color}
\usepackage{bm}
\usepackage[colorlinks, linkcolor=red]{hyperref}
\usepackage{threeparttable}
\usepackage[ruled]{algorithm}
\usepackage[american]{babel}
\usepackage{microtype}
\begin{document}
\title{Attention Mechanism Enhanced Kernel Prediction Networks for Denoising of Burst Images}
\name{Bin Zhang$^{1}$, Shenyao Jin$^{2}$, Yili Xia$^{1}$, Yongming Huang$^{1}$, and Zixiang Xiong$^{3}$}
\address{$^{1}$School of Information Science and Engineering, Southeast University, Nanjing 210096, P. R. China\\
$^{2}$Jiangsu Industrial Technology Research Institute, Nanjing 211111, P. R. China\\
$^{3}$Department of Electrical and Computer Systems Engineering, Texas A\&M University, TX 77843, USA\\Emails: \{220180920, yili\_xia, huangym\}@seu.edu.cn, cas\_colinjin@163.com, zx@ece.tamu.edu}
%
\maketitle
\vspace{-8pt}
\begin{abstract}
\vspace{-4pt}
Deep learning based image denoising methods have been extensively investigated. In this paper, attention mechanism enhanced kernel prediction networks (AME-KPNs) are proposed for burst image denoising, in which, nearly cost-free attention modules are adopted to first refine the feature maps and to further make a full use of the inter-frame and intra-frame redundancies within the whole image burst. The proposed AME-KPNs output per-pixel spatially-adaptive kernels, residual maps and corresponding weight maps, in which, the predicted kernels roughly restore clean pixels at their corresponding locations via an adaptive convolution operation, and subsequently, residuals are weighted and summed to compensate the limited receptive field of predicted kernels. Simulations and real-world experiments are conducted to illustrate the robustness of the proposed AME-KPNs in burst image denoising.
\end{abstract}

%
\vspace{-8pt}
\section{Introduction}
\vspace{-8pt}

Image denoising is a longstanding prerequisite problem due to limitations of the image capture system which is inherently corrupted by the measurement noise.  Over the past decades, numerous image denoising solutions have been proposed \cite{Buades2005NLM, Dabov2008BM3D, Lebrun2012An, Maggioni2012VBM4D}, and the key difference among them lies in how to determine an efficient scheme to select suitable pixels and to compute the averaging weights. For example, BM3D and its variants find relative pixels by block matching and averaging these pixels via empirical Wiener filters \cite{Dabov2008BM3D, Lebrun2012An, Maggioni2012VBM4D}. However, their performances are restricted to specific forms of prior, and many hand-tuned parameters are involved in the training process, which makes them not always operate well in complicated scenarios.

The success of deep learning has yielded many neural network based approaches for image denoising and other fundamental vision tasks~\cite{Kim2019GRDN, Wang2019EDVR, Liu2019Learning}. The deep convolution neural network in~\cite{Zhang2016Beyond} is one of the first attempts for image denoising, and it provides an end-to-end denoising scheme for a single frame by adopting a residual learning strategy to estimate the noise existence within the input image. In~\cite{Mao2016RED}, residual encoder-decoder networks with skip connections are proposed, in which, convolution layers act as the feature extractor to capture abstraction of noisy image contents, and subsequently, image details are recovered by deconvolution layers. Furthermore, an extra estimation subnetwork is utilized in \cite{Guo2019CBDNet} to generate a noise map so as to improve the generalization ability of residual encoder-decoder networks to real-world images. More recently, the effectiveness of the attention mechanism on the blind image denoising has been discussed in~\cite{Anwar2019RIDNet}, in which a feature attention module is adopted to calibrate the weights associated to important features within maps.

Except for the above single-frame studies, there are also multi-frame image denoising frameworks based on neural networks, and in general, they bring improvements as compared to single-frame ones, due to their capability to use temporal coherence within the image burst \cite{Godard2018dDeep,Zhao2019RFCN,Mildenhall2018KPN,Marinc2019MKPN,Xu2019Deformable,Niklaus2017AdaptiveConv}. Among them, kernel-based methods, such as kernel prediction networks (KPNs) and multi-KPN (MKPNs) \cite{Mildenhall2018KPN,Marinc2019MKPN,Xu2019Deformable}, have drawn a significant amount of attention. The main idea behind them is to predict per-pixel spatially-adaptive kernels, which are further involved in an adaptive convolution operation to denoise burst images \cite{Niklaus2017AdaptiveConv}. However, their performances are limited by the receptive field of per-pixel kernels.



To address this issue, attention mechanism enhanced KPNs (AME-KPNs) based on a modified U-Net architecture \cite{Ronneberger2015UNet} are proposed for denoising of burst images. Although their structure is similar to the original KPNs, they operate in a fundamentally different way. In particular, the attention mechanism is employed to refine feature maps at multiple scales. In this sense, the spatial-temporal relation within the burst of images can be well preserved by the proposed networks. Besides, not only per-pixel spatially-adaptive kernels are predicted, but also residuals are produced to calibrate the coarse prediction of the denoised image, so as to enlarge the receptive field of the adaptive convolution. Extensive simulations on both synthetic and real-world data are conducted to verify the effectiveness of the proposed AME-KPNs.
%
%
\vspace{-10pt}
\section{Proposed AME-KPNs}
\vspace{-8pt}
For a given noisy burst of $N$ images, $\mathbf{X} = [X_1, X_2, \ldots, X_N]$, where $[\cdot, \cdot]$ is the concatenation operation along the channel dimension
and $X_1$ is selected as the reference, our goal is to denoise $X_1$ with the help of neighboring frames, $X_2, \ldots, X_N$, and to yield a denoised image $\hat{Y}$.
\begin{figure*}[ht!]
	\vspace{-16pt}
	\centering
	\includegraphics[width=0.8\linewidth]{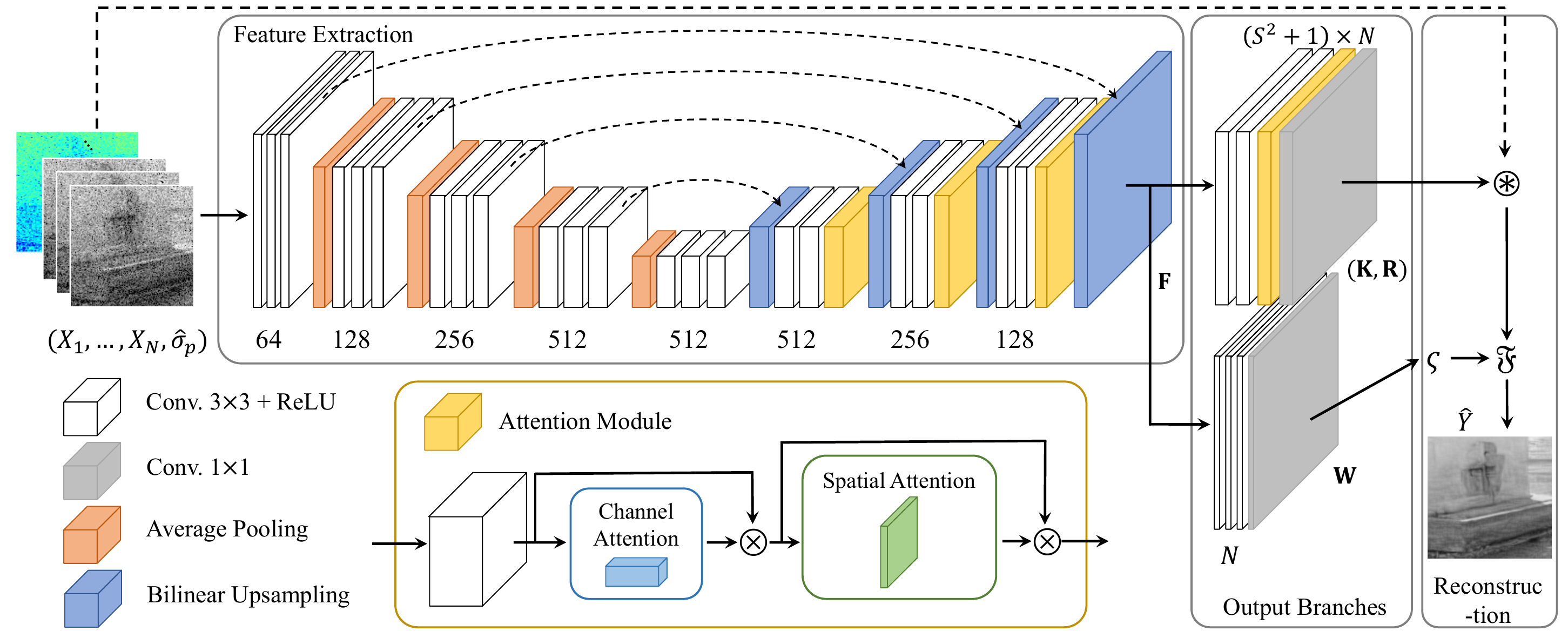}
	\caption{The architecture of our proposed AME-KPNs, in which $\circledast$ and $\otimes$ denote the adaptive convolution and element-wise multiplication operators, respectively. The number of channels of the output tensor  is denoted either below or above the corresponding layer.}
	\label{fig:arch}
\end{figure*}
%
\vspace{-8pt}
\subsection{Network Architecture}
\vspace{-4pt}
Fig. \ref{fig:arch} shows the architecture of the proposed AME-KPNs, which is composed of three main modules, that is, feature extraction, output branches, and reconstruction. The module of feature extraction, termed as $M_e$, is a U-Net like structure that is an encoder-decoder framework with skip connections, where the encoder sequentially transforms input frames into lower-resolution feature embeddings and the decoder correspondingly expands the features back for a full-resolution estimation~\cite{Ronneberger2015UNet}. The skip connections, denoted by dash lines in Fig. \ref{fig:arch}, are added to concatenate congruent layers in encoder and decoder to use both low-level and high-level features for the estimation process. Given an estimated noise map of the reference image $X_1$, that is, $\hat{\sigma}_p$, which is also taken as one of the inputs for a better generalization ability, the feature map $\mathbf{F}$ can be extracted by the backbone network $M_e$ as
\vspace{-4pt}
\begin{equation}
	\mathbf{F} = M_e([\mathbf{X}, \hat{\sigma}_p]).
	\vspace{-4pt}
\end{equation}
Subsequently, it is fed into two output branches, that is, $M_{kr}$ and $M_w$. As a result, per-pixel spatially-adaptive kernels, $\mathbf{K} \!=\! [K_1, K_2, \ldots, K_N]$, residuals, $\mathbf{R} \!=\! [R_1, R_2, \ldots, R_N]$, and corresponding per-pixel weight maps, $\mathbf{W} \!=\! [W_1, W_2, \ldots, W_N]$, are obtained as
\vspace{-4pt}
\begin{align}
	[\mathbf{K}, \mathbf{R}] &= M_{kr}(\mathbf{F})\label{eq:out_braches} \\
	\mathbf{W} &= M_w(\mathbf{F}),
	\vspace{-8pt}
\end{align}
in which, both $\mathbf{K}$ and $\mathbf{R}$ are the output of $M_{kr}$, and $\mathbf{R}$ is obtained by splitting the last $N$ channels.

Because several extra calculations are involved to predict the denoised image, the proposed AME-KPNs operate in a different way as compared with the original KPNs. Specifically, for each  spatially-adaptive kernel of size $S \!\times\!S$, the convolved value is a linear combination of the input noisy pixel and its square neighboring pixels via an inner product operation~\cite{Niklaus2017AdaptiveConv}. The kernel filters the information of importance for restoring the original clean image and excludes other noisy pixels. In this way, the convolved value is related to the surrounding pixels within the $S \!\times\! S$ region only, and hence, the efficiency of spatially-adaptive kernels is limited by their sizes. A small kernel is capable of preserving edges and details of images and fixing small misalignments, but is powerless for a large area. In contrast, a large kernel unexpectedly oversmoothes the desired edges and details~\cite{Marinc2019MKPN}. Next, residual maps, which are created from the feature map $\mathbf{F}$ according to \eqref{eq:out_braches}, compensate the coarse output of the adaptive convolution; this is beneficial for handling the large misalignment, due to the large receptive field, and for preserving details at the same time. Finally, the denoised image $\hat{Y}$ can be obtained through the reconstruction module, given by
%
%
\vspace{-4pt}
\begin{align}
	\hat{Y} &= \frac{1}{N} \sum_{i=1}^{N} \hat{Y}_i
	= \frac{1}{N}\sum_{i=1}^{N} \mathfrak{F} \left( X_i \!\circledast\! K_i, R_i, \varsigma(W_i)\right) \nonumber\\
	&= \frac{1}{N} \sum_{i=1}^{N} \varsigma(W_i)\!\cdot\! \left(X_i \!\circledast\! K_i\right) \!+\! \left(1 \!-\! \varsigma(W_i)\right) \!\cdot\! R_i,
	\vspace{-10pt}
\end{align}
where $\hat{Y}_i$ is the predicted image which corresponds to $X_i$, $\mathfrak{F}(\cdot, \cdot, \cdot)$ represents the reconstruction module, and $\varsigma(\cdot)$ is the sigmoid function to restrict the values of weight maps within the range $[0, 1]$.

\vspace{-8pt}
\subsection{Attention Mechanism}\label{sec:attention_mechanism}
\vspace{-4pt}
The burst of images are usually continuous frames, and hence, feature strong temporal redundancies. However, these images may contain motions from both the hand shake and the object movement, which make neighbor frames not equally informative to the reference one, and result in an adverse effect on the denoising process. Therefore, how to make an efficient use of both the inter-frame and intra-frame spatial relations within these frames is a critical issue.

The proposed AME-KPNs alleviate the above problem by adopting the attention mechanism, which dynamically aggregates the information of neighboring frames at the pixel level and assigns aggregated weights for each pixel within the feature map. Specifically, convolution layers with two attention modules, that is, channel attention $M_{Ac}$ and spatial attention $M_{As}$, are exploited at each block of the decoder with multiple receptive fields at multiple layers,
as shown in Fig.~\ref{fig:arch}. Similar to~\cite{Woo2018CBAM}, two attention maps are sequentially generated and then multiplied to the input feature map for adaptive feature refinement along two separate dimensions. Given the input feature $\mathbf{F}_d$, we have,
\vspace{-4pt}
\begin{align}
	\mathbf{F}_c = M_{Ac}(\mathbf{F}_d) \otimes \mathbf{F}_d,\\
	\mathbf{F}_s = M_{As}(\mathbf{F}_c) \otimes \mathbf{F}_c,
	\vspace{-8pt}
\end{align}
in which $\mathbf{F}_c$ and $\mathbf{F}_s$ are feature maps refined by the channel attention module and the spatial attention one, respectively. In this way, both the inter-frame and intra-frame redundancies within burst images are implicitly taken into the subsequent part of the proposed networks.
%
\vspace{-8pt}
\subsection{Loss Function}
\vspace{-4pt}
Similar to~\cite{Mildenhall2018KPN, Marinc2019MKPN, Xu2019Deformable}, the proposed AME-KPNs are trained in the linear raw domain, and the linear output $\hat{Y}$ is further performed a gamma correction to generate the final denoised image for a better perceptual quality. Given the ground truth image $Y_{gt}$, the basic loss function for this aim is given by
\vspace{-4pt}
\begin{equation}
	\ell(\hat{Y}, Y_{gt}) \!=\! ||\Gamma(\hat{Y}) \!-\! \Gamma(Y_{gt})||_2^2 \!+\! \lambda ||\nabla\Gamma(\hat{Y}) \!-\! \nabla\Gamma(Y_{gt})||_1,
\label{eq:cost1}
	\vspace{-4pt}
\end{equation}
where $\Gamma(\cdot)$ is the gamma correction function and $\nabla$ denotes the gradient operator. The second term on the right hand side of \eqref{eq:cost1} is a regularizer to constrain the smoothness of $\hat{Y}$, and $\lambda$ represents the regularization coefficient. To avoid the situation where $\ell(\hat{Y}, Y_{gt})$ converges to a local minimum, an annealed loss is further adopted to yield the total loss function $\mathcal{L}(\hat{Y}, Y_{gt})$, given by
\vspace{-4pt}
\begin{equation}
	\mathcal{L}(\hat{Y}, Y_{gt}) = \ell(\hat{Y}, Y_{gt}) + \beta \alpha^t \sum_{i=1}^{N} \ell(\hat{Y}_i, Y_{gt}),
	\vspace{-8pt}
\end{equation}
in which, $\alpha$ and $\beta$ are hyperparameters, typically, $\alpha=0.9998$ and $\beta=100$, and $t$ is the number of iterations used in the optimization process~\cite{Mildenhall2018KPN,Xu2019Deformable}. Note that, when $\beta \alpha^t \!>>\! 1$, the neighboring frames are shifted and aligned to the reference, and all of them are denoised independently, since the value of the annealed term $\beta \alpha^t \sum_{i=1}^{N} \ell(\hat{Y}_i, Y_{gt})$ is always prominent. As the iteration evolves, this constraint disappears and the denoised image $\hat{Y}$ is obtained based on the entire burst.
%
\vspace{-10pt}
\section{Simulations}
\vspace{-8pt}
Experiments were conducted to qualitatively and quantitatively evaluate the superiority of the proposed AME-KPNs over the original KPNs \cite{Mildenhall2018KPN} and MKPNs \cite{Marinc2019MKPN} for denoising of burst images. Both grayscale and color images were considered, with the latter ones handled in terms of the R, G, B channels independently by using the considered networks.
\vspace{-8pt}
\subsection{Synthetic Dataset}\label{sec:synthetic}
\vspace{-4pt}
We followed the same procedure in \cite{Mildenhall2018KPN,Marinc2019MKPN} to synthesize the training data and to estimate the noise map $\hat{\sigma}_p$. The reference patches of size $512\!\times \!512$ were sampled from the images in the Adobe-5K dataset \cite{Bychkovsky2011FiveK} to create the ground truth and the remaining $N-1$ neighbor images. In order to simulate the misalignment between consecutive frames caused by motions, the neighbor images were offset from the reference by $x_i$ and $y_i$ for $i=1,\ldots,N-1$, where $x_i$ and $y_i$ are offsets of the $i$th image in horizontal and vertical directions, respectively. Their values were sampled with probability $n/N$ from a 2D uniform integer distribution between $[-64,64]$, otherwise from a 2D uniform integer distribution between $[-8,8]$, where $n\!\sim\! \text{Poisson}(\lambda)$ and $\lambda\!=\! 1.5$. Then, the burst images were $4\times$ downsampled in each dimension by using a box filter to reduce compression artifacts. The noise model of camera sensors in \cite{Mildenhall2018KPN} was considered to build up noisy burst images. The noisy measurement $x_p$ of pixel $p$ in each image was Gaussian distributed with mean $y_p$, which is the true intensity of pixel $p$, and variance $\sigma_r^2+\sigma_s y_p$, so that, $x_p\!\sim\! \mathcal{N}(y_p, \sigma_r^2+\sigma_s y_p)$, where read and shot noise parameters $\sigma_r$ and $\sigma_s$ were sampled uniformly from $[10^{-3}, 10^{-1.5}]$ and $[10^{-4}, 10^{-2}]$, respectively. Finally, due to the non-blind denoising nature of the considered networks, the noise map $\hat{\sigma}_p$ of the reference was estimated as $\hat{\sigma}_p \!=\! \sqrt{\sigma_r^2 + \sigma_s \max(x_p, 0)}$, which is also taken as the input of networks.
%
%
%
%
\vspace{-8pt}
\subsection{Training Details}
\vspace{-4pt}
All the networks were implemented in PyTorch\footnote{The source codes implemented in PyTorch are available at \href{https://github.com/z-bingo/Attention-Mechanism-Enhanced-KPN}{https://github.com/z-bingo/Attention-Mechanism-Enhanced-KPN}.} and optimized by Adam with an initial learning rate $10^{-4}$. The learning rate was decreased by a factor of $10^{-0.05} (\approx\! 0.8913)$ per epoch until it reached $5\!\times\! 10^{-6}$ \cite{Mildenhall2018KPN,Marinc2019MKPN}. All the networks were trained $8\!\times\! 10^4$ iterations with a batch size of $16$ bursts composed of $N=8$ images, and it took roughly $8$ hours for each network on an NVIDIA GTX 1080Ti GPU.
\vspace{-8pt}
\subsection{Evaluation on Synthetic Dataset and Real-World Images}
\begin{figure}[!t]
	\vspace{-8pt}
	\centering
	\subfloat[]{
		\begin{minipage}[t]{0.16\linewidth}
			\centering
			\includegraphics[width=\linewidth]{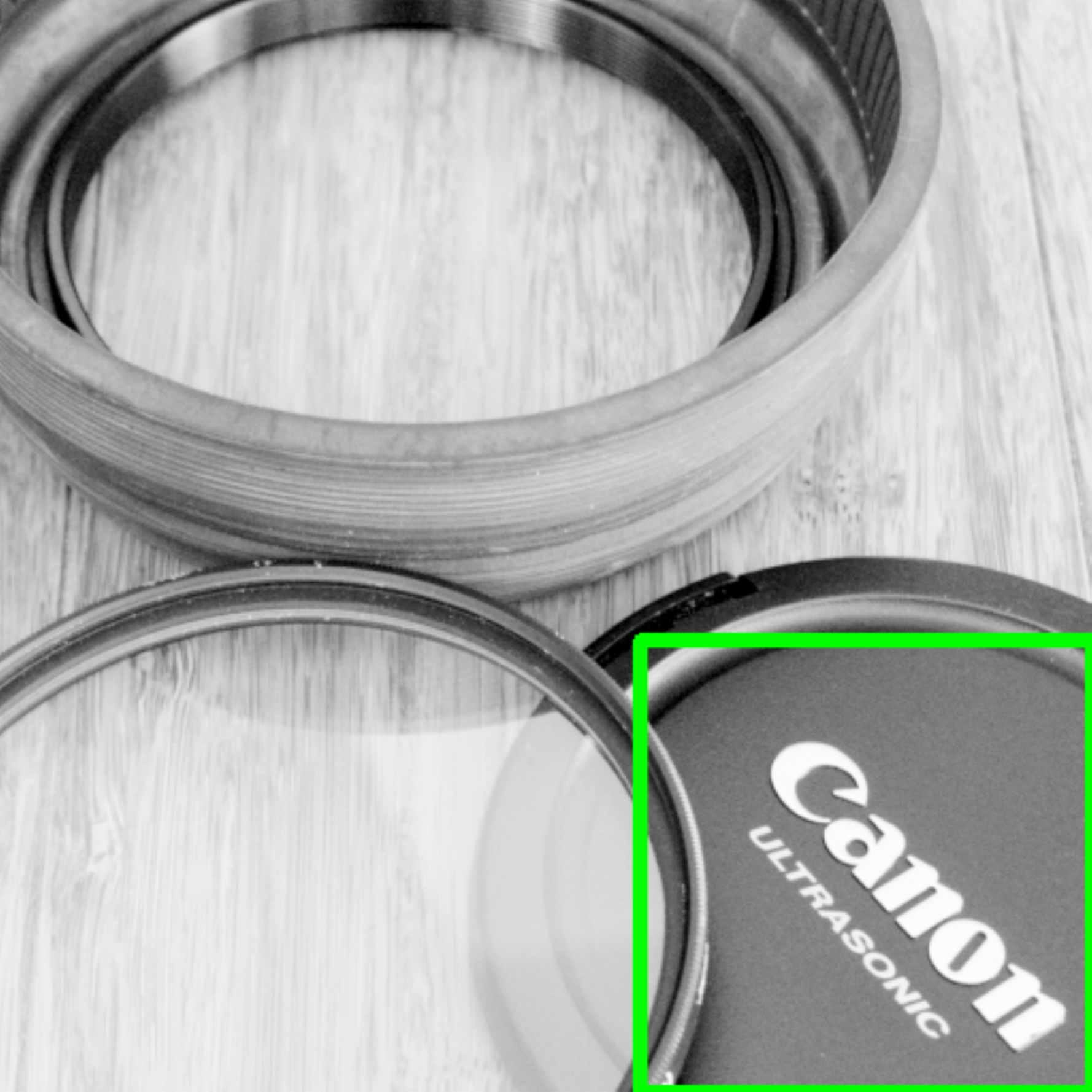}
			\includegraphics[width=\linewidth]{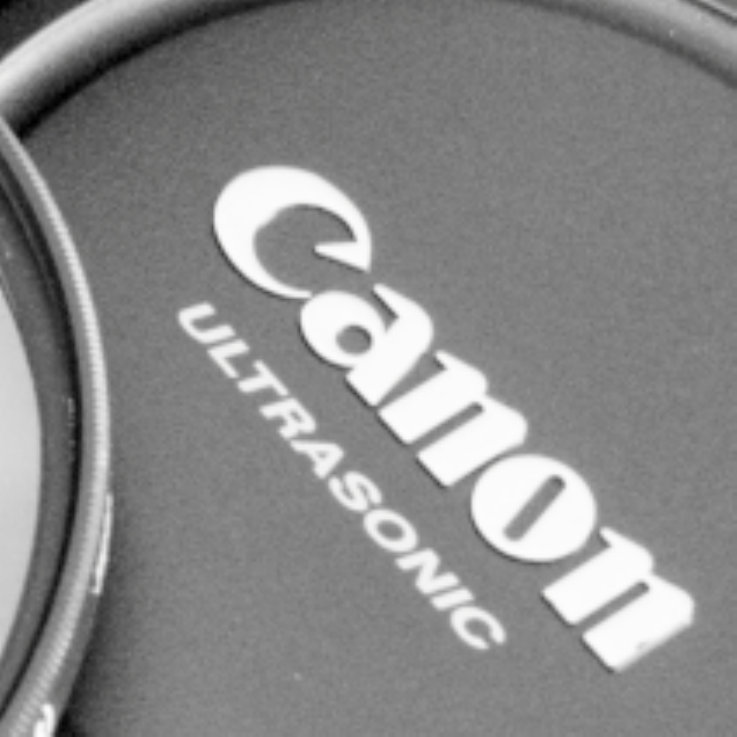}
		\end{minipage}
	}
	\hspace{-6pt}
	\subfloat[]{
		\begin{minipage}[t]{0.16\linewidth}
			\centering
			\includegraphics[width=\linewidth]{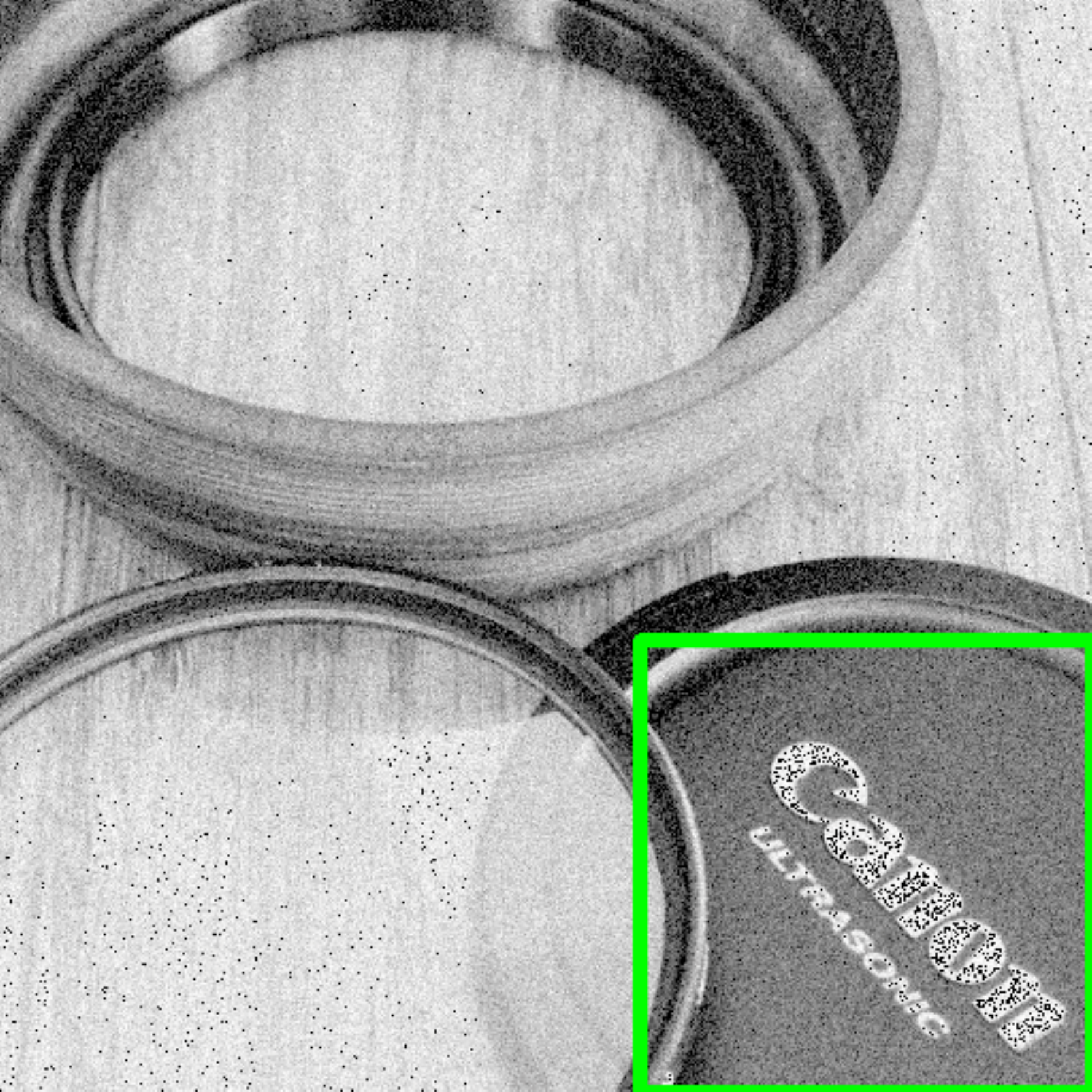}
			\includegraphics[width=\linewidth]{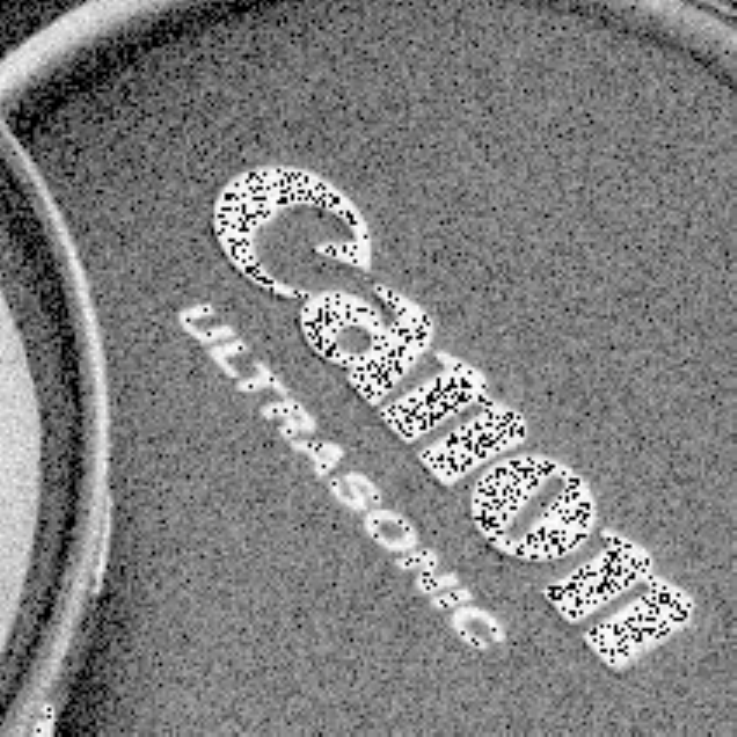}
		\end{minipage}
	}
	\hspace{-6pt}
	\subfloat[]{
		\begin{minipage}[t]{0.16\linewidth}
			\centering
			\includegraphics[width=\linewidth]{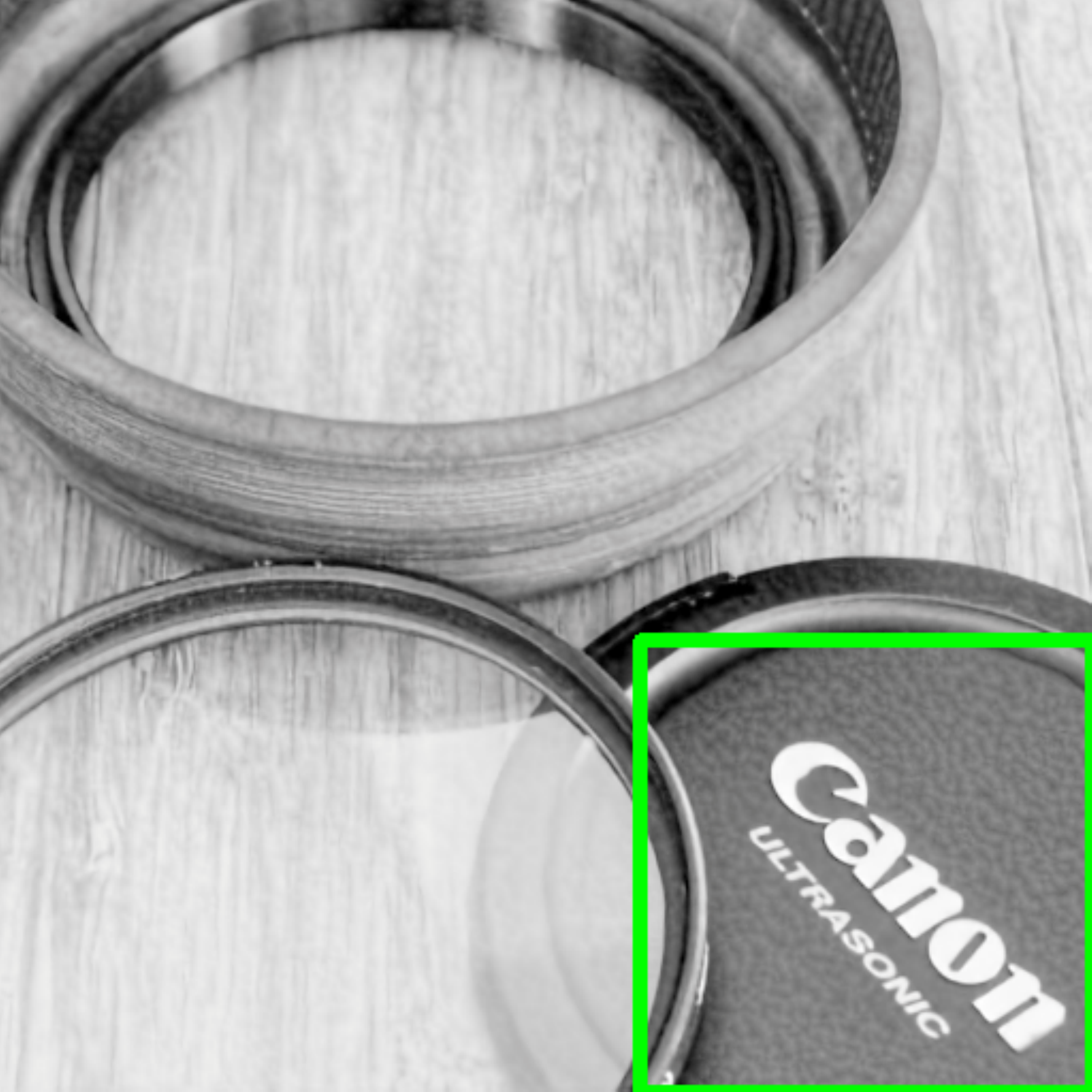}
			\includegraphics[width=\linewidth]{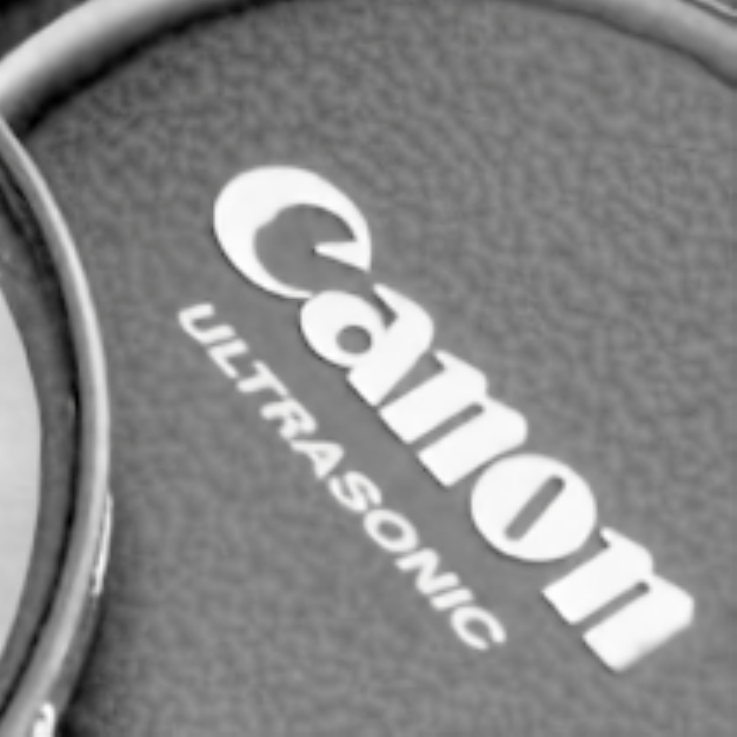}
		\end{minipage}
	}
	\hspace{-6pt}
	\subfloat[]{
		\begin{minipage}[t]{0.16\linewidth}
			\centering
			\includegraphics[width=\linewidth]{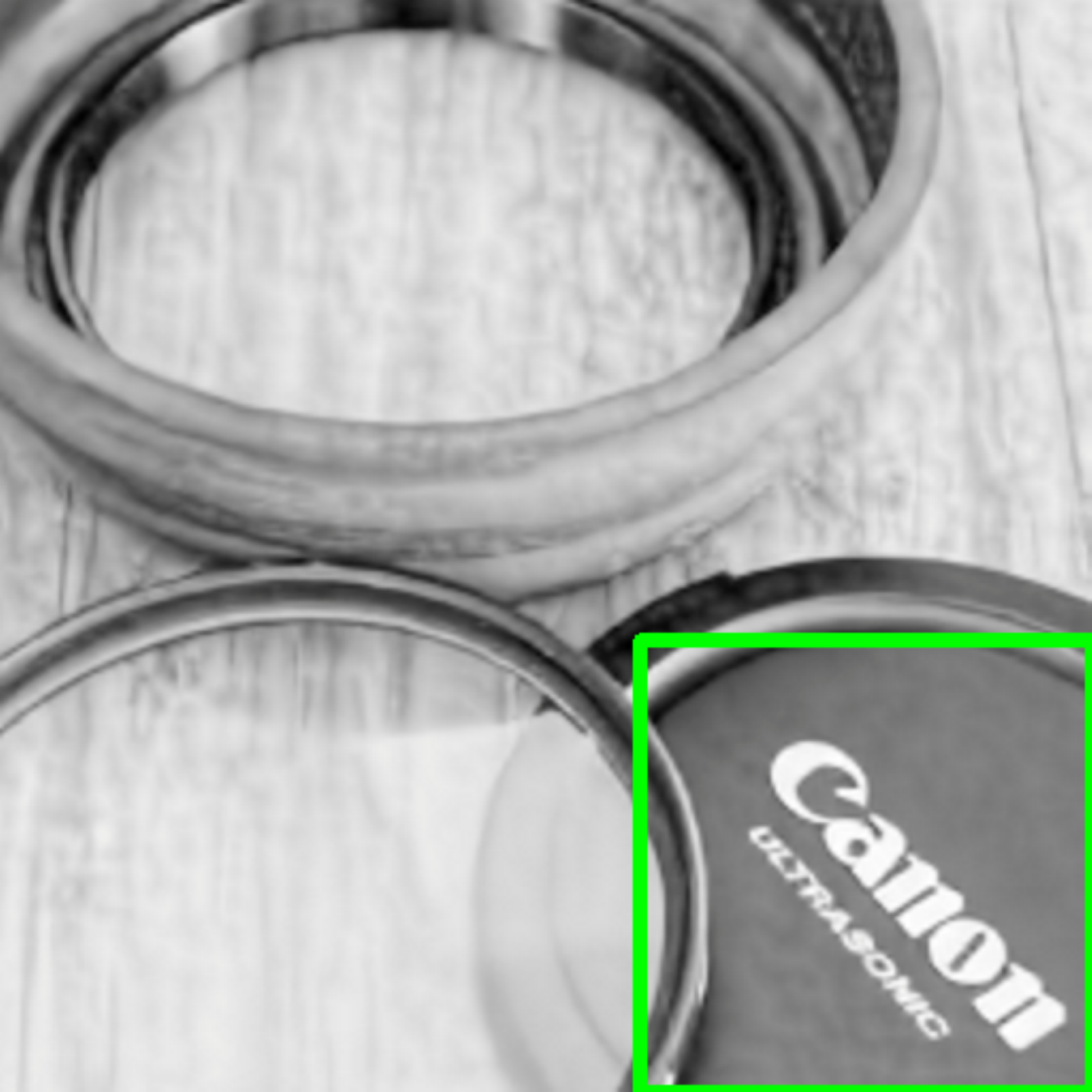}
			\includegraphics[width=\linewidth]{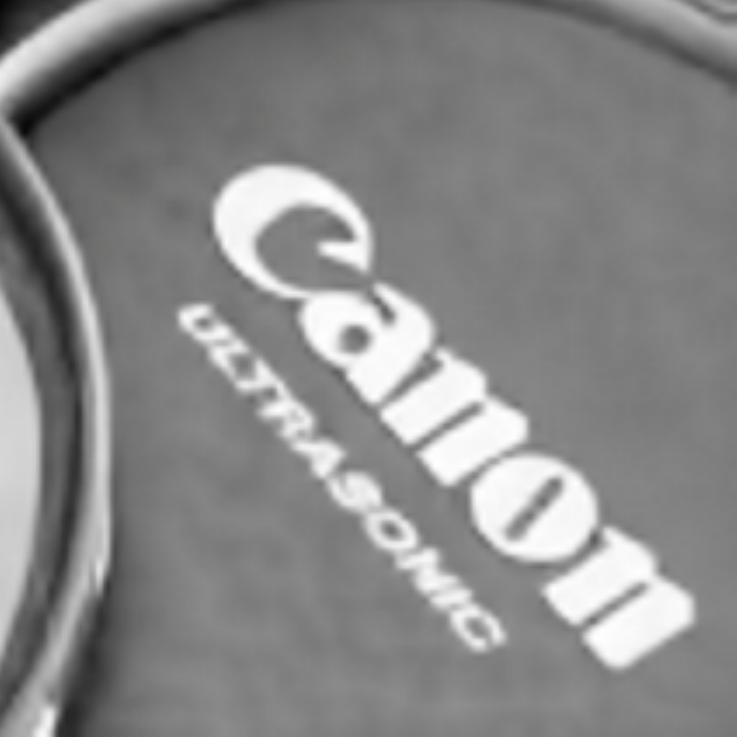}
		\end{minipage}
	}
	\hspace{-6pt}
	\subfloat[]{
		\begin{minipage}[t]{0.16\linewidth}
			\centering
			\includegraphics[width=\linewidth]{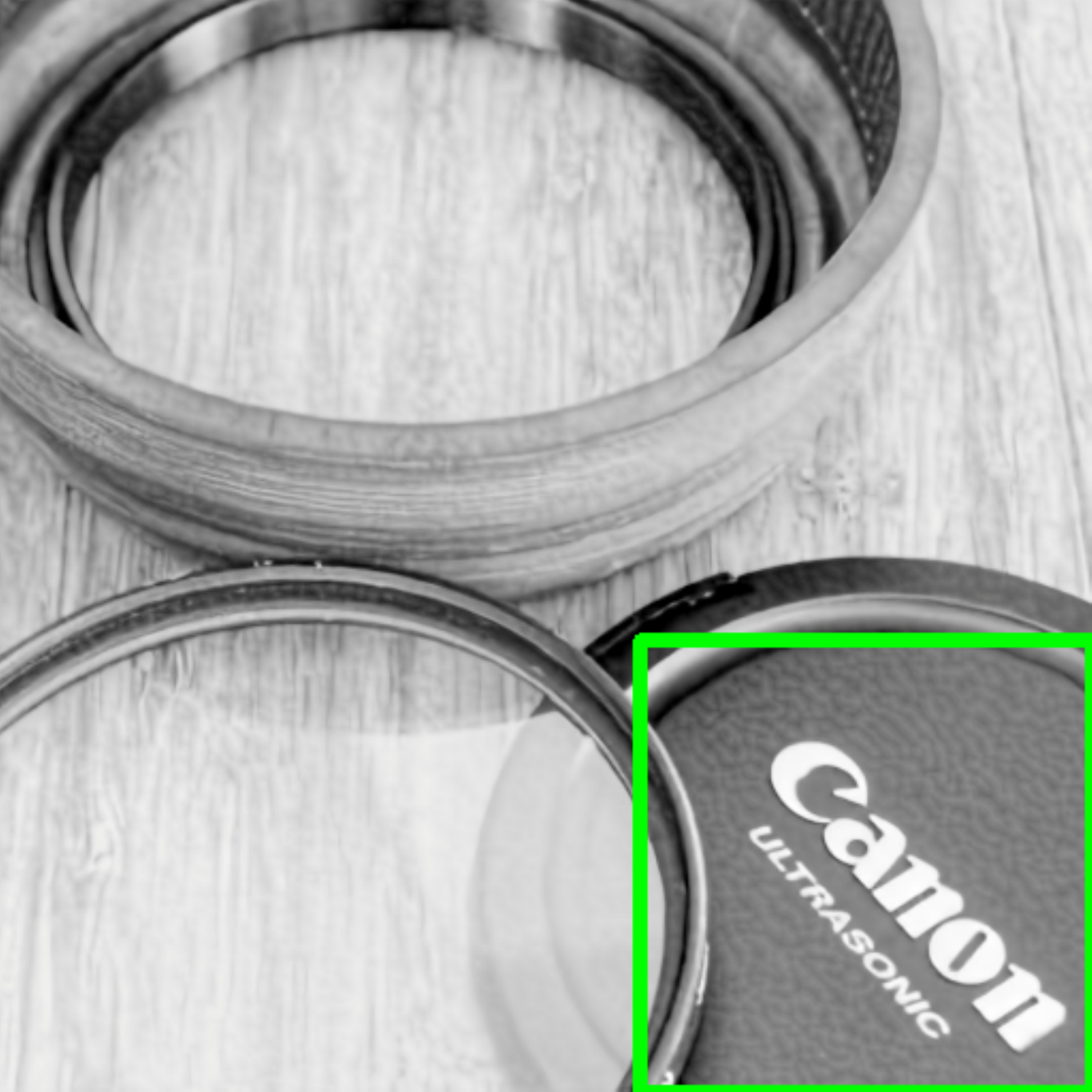}
			\includegraphics[width=\linewidth]{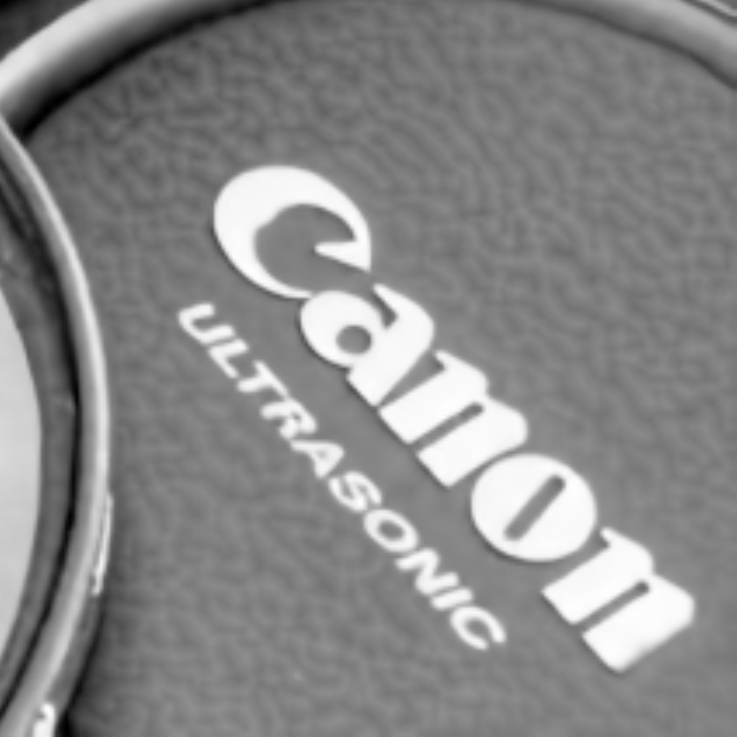}
		\end{minipage}
	}
	\hspace{-6pt}
	\subfloat[]{
		\begin{minipage}[t]{0.16\linewidth}
			\centering
			\includegraphics[width=\linewidth]{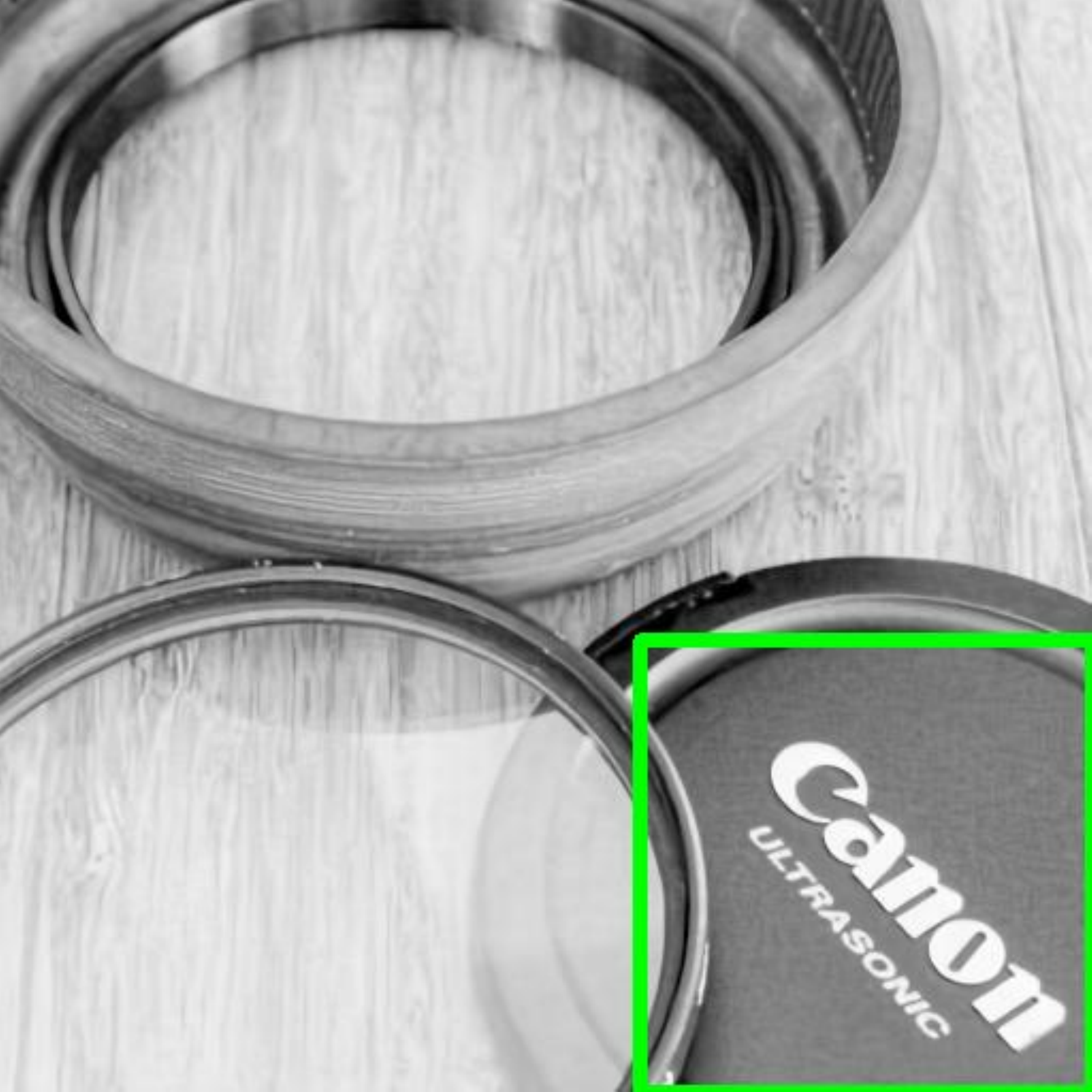}
			\includegraphics[width=\linewidth]{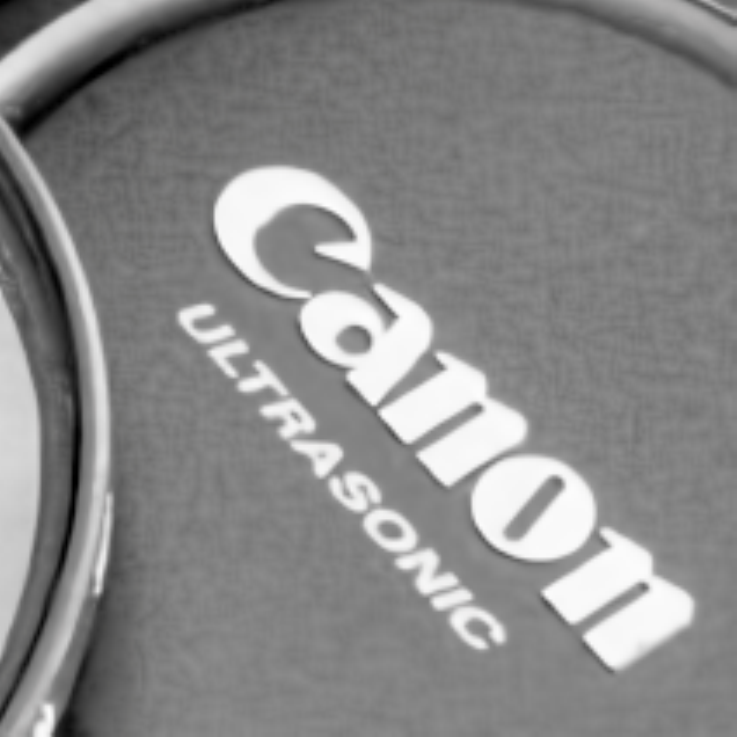}
		\end{minipage}
	}
	\hspace{-6pt}
	\caption{An illustration of denoised images at Gain 3. Images from (a) to (f) are the ground truth, the noised reference and the results of KPNs ($5\times5$), KPNs ($21\times21$), MKPNs and the proposed AME-KPNs ($5\times5$), respectively.}
	\label{fig:eval_syn}
\end{figure}

\vspace{-4pt}
In the first stage, we compared the proposed AME-KPNs with $5\!\times\!5$ kernels against KPNs with $5\!\times\!5$ kernels and $21 \!\times\! 21$ kernels, and MKPNs of different kernel sizes $S \!\in\! \left\{1,3,5,7,9\right\}$ on the synthetic dataset and against different levels of noise. Especially, the noise level in Gain 4 was far beyond the training range. Table \ref{tab:quan_syn} shows the superiority of our proposed AME-KPNs over KPNs and MKPNs in terms of both the peak signal to noise ratio (PSNR) in dB  and the structural similarity index (SSIM). An illustration of the conducted simulations is presented in Fig.~\ref{fig:eval_syn}, where several significant differences among the denoised results of different models can be observed, especially on the cap of the camera len, zoomed in the green box. As shown in Fig.~\ref{fig:eval_syn}(c), obvious long-grained artifacts were encountered by KPNs with small $5\!\times\!5$ kernels in flat areas within the image. On the other hand, large $21\!\times\!21$ kernels in KPNs brought about unclear details, which can be seen from the letters ``ULTRASONIC" in Fig.~\ref{fig:eval_syn}(d). In Fig.~\ref{fig:eval_syn}(e), although MKPNs made a trade-off in denoising quality between smooth areas and clear details by producing multiple sizes of kernels for each pixel, its overall denoising performance was still worse than that of the proposed AME-KPNs, as shown in Fig.~\ref{fig:eval_syn}(f).
\begin{table}[!t]
	\vspace{-8pt}
	\centering
	\caption{Quantitative comparison among different networks on the synthetic dataset at various gains (noise levels) \tnote{$\ast$}.}
	\label{tab:quan_syn}
	\resizebox{\linewidth}{!}{
		\begin{threeparttable}
			\begin{tabular}{ccccccccc}
				\toprule[1pt]
				\multicolumn{1}{c}{\multirow{2}{*}{Algorithms}} & \multicolumn{2}{c}{Gain 1} & \multicolumn{2}{c}{Gain 2} & \multicolumn{2}{c}{Gain 3} & \multicolumn{2}{c}{Gain 4} \\
				\multicolumn{1}{c}{}                            & PSNR         & SSIM         & PSNR         & SSIM         & PSNR         & SSIM         & PSNR         & SSIM         \\
				\midrule[0.5pt]
				Reference frame									& 33.98  	   & 0.8625 	  &	26.94		 & 0.6286		& 21.50 	   & 0.3980 	  &	16.44		 & 0.2027		\\
				KPN, $5\times 5$                                & 37.94        & 0.9767       & 35.75        & 0.9524       & 33.70        & 0.9223       & 30.99        & 0.8603       \\
				KPN, $21\times 21$                              & 37.17        & 0.9754       & 35.87        & 0.9523       & 33.47        & 0.9204       & 30.61        & 0.8573        \\
				MKPN                     & 38.29        & 0.9719       & 35.82        & 0.9489       & 35.43        & 0.9122       & 30.66        & 0.8586       \\
	            AME-KPN, $5\times 5$                         & \bf{39.92}   & \bf{0.9776}  & \bf{37.02}   & \bf{0.9561}  & \bf{35.46}   & \bf{0.9295}  & \bf{32.29}   & \bf{0.8663}      \\
				\bottomrule[1pt]
			\end{tabular}
			\begin{tablenotes}
				\item[$\ast$] Parameters of the four gains considered. Gain 1: $\sigma_r=5\times10^{-3}, \sigma_s=10^{-3}$, Gain 2: $\sigma_r=2\times 10^{-2}, \sigma_s=4.3\times10^{-3}$, Gain 3: $\sigma_r=5\times10^{-2}, \sigma_s=10^{-2}$, and Gain 4: $\sigma_r=8\times10^{-2}, \sigma_s = 2.3\times10^{-2}$.
			\end{tablenotes}
			
		\end{threeparttable}
	}
\end{table}

\begin{figure}[!t]
	\vspace{-16pt}
	\centering
	\subfloat[Noisy refrence]{
		\includegraphics[width=0.48\linewidth]{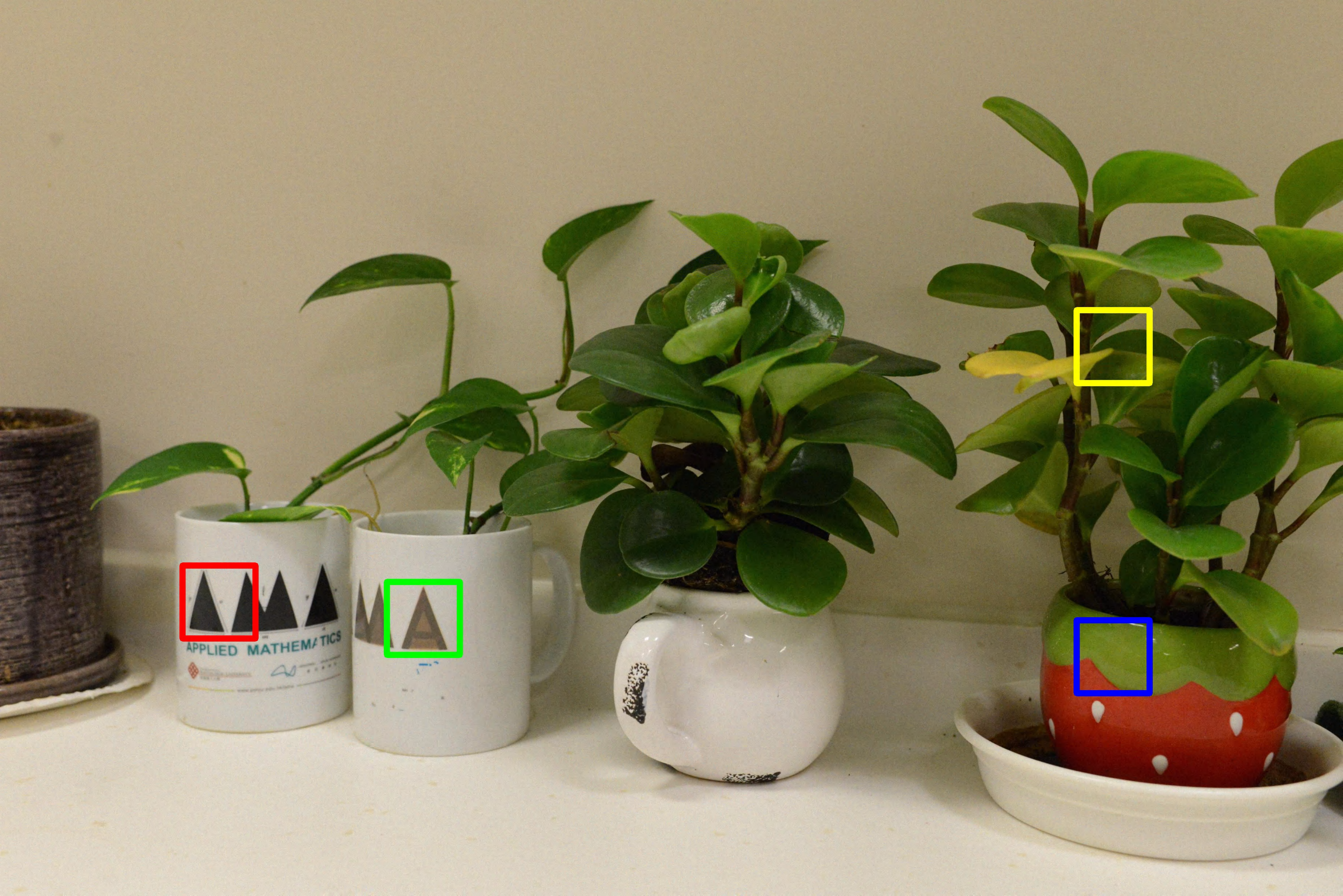}
	}
	\hspace{-6pt}
	\subfloat[AME-KPN output]{
		\includegraphics[width=0.48\linewidth]{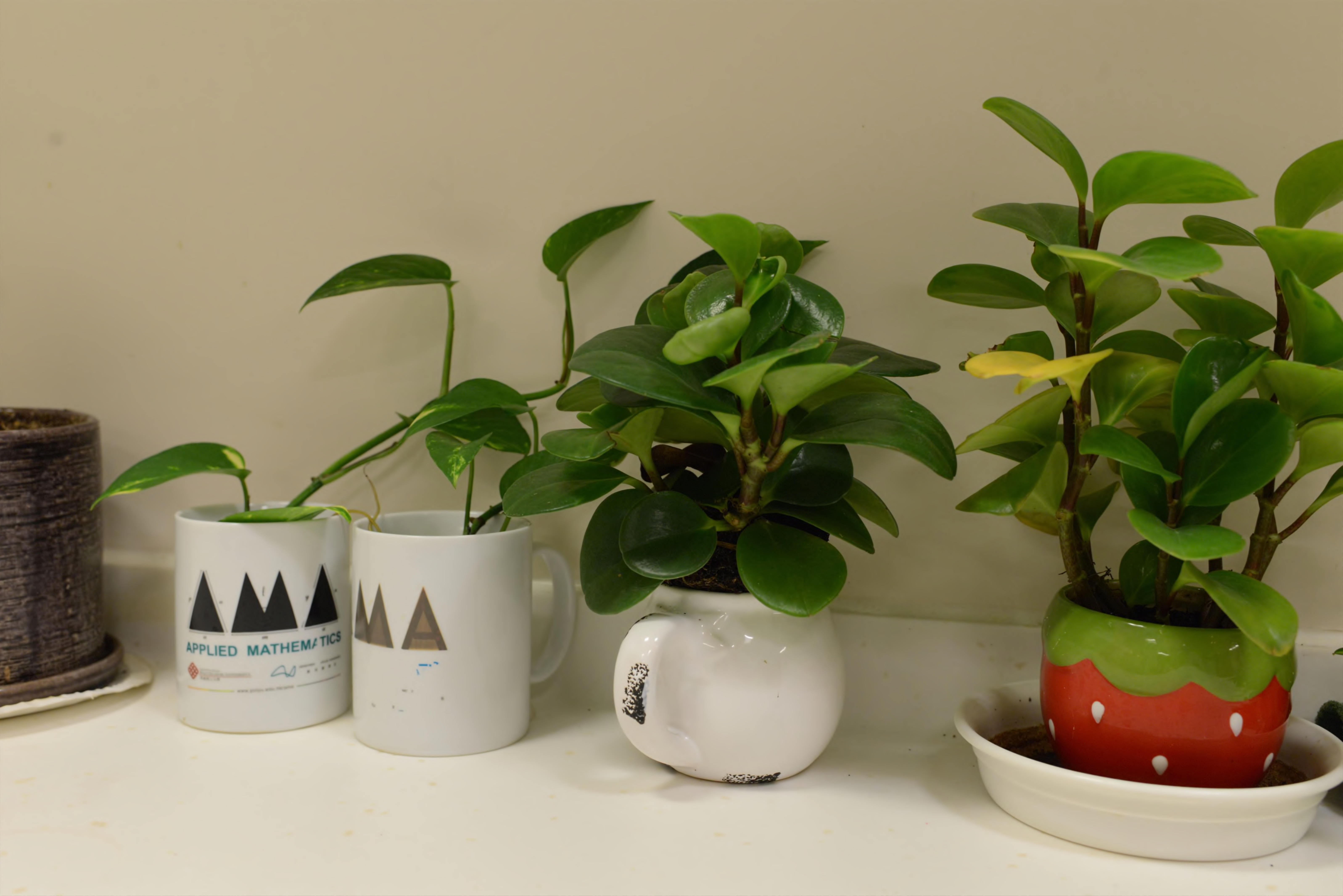}
	}
	\hspace{-6pt}
	\vfill
	\centering
	\vspace{-8pt}
	\subfloat[]{
		\includegraphics[height=0.48\linewidth]{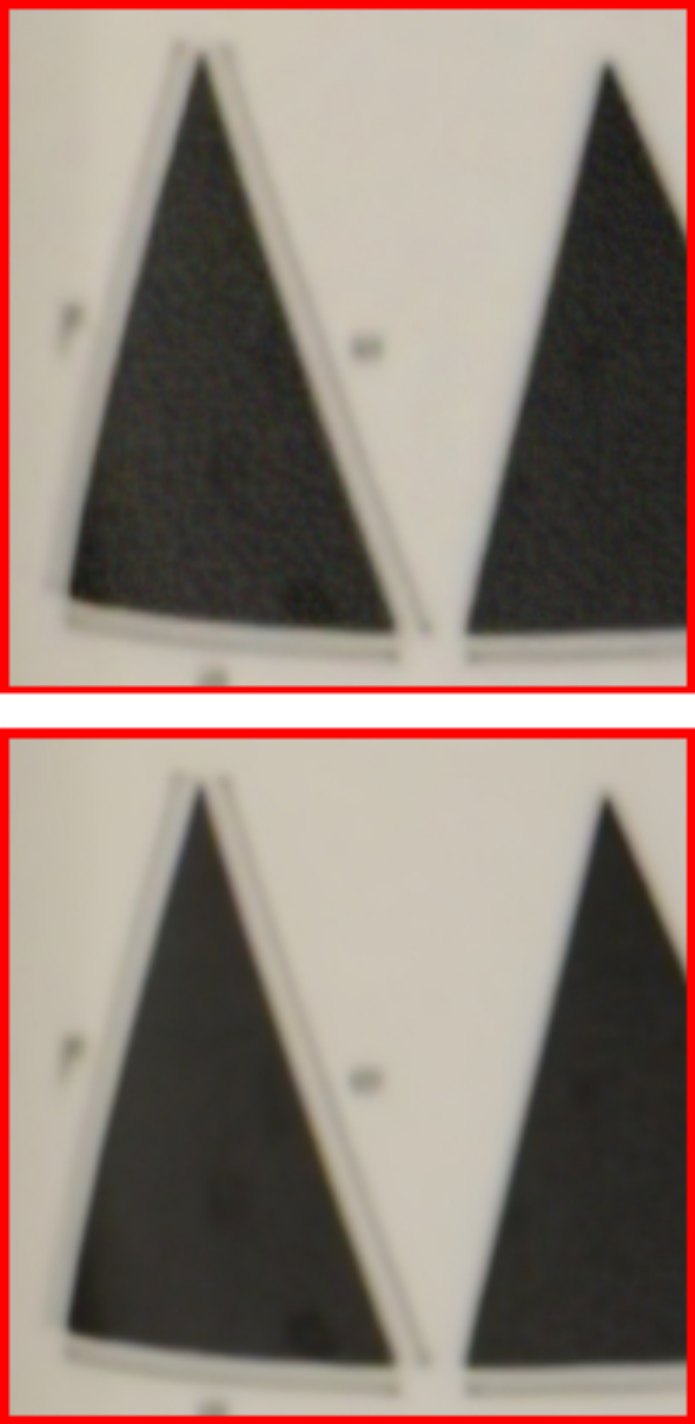}
	}
	\hspace{-6pt}
	\subfloat[]{
		\includegraphics[height=0.48\linewidth]{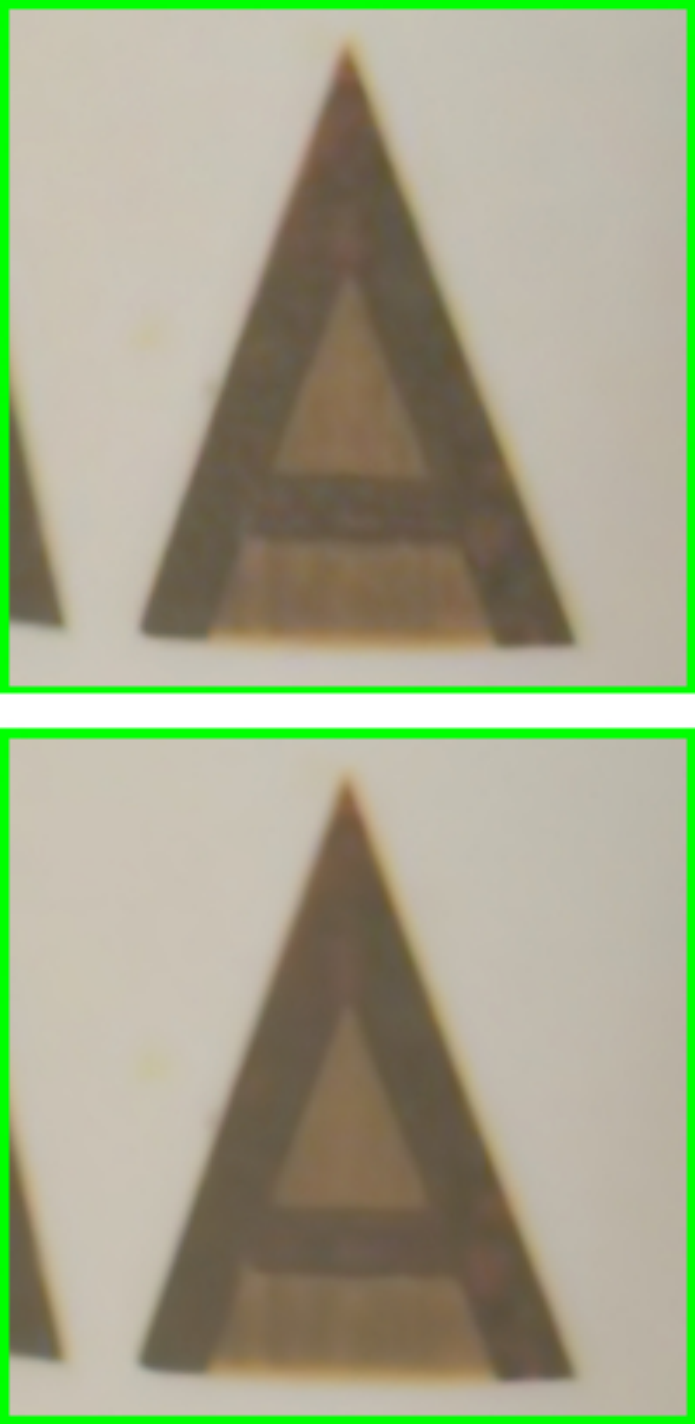}
	}
	\hspace{-6pt}
	\subfloat[]{
		\includegraphics[height=0.48\linewidth]{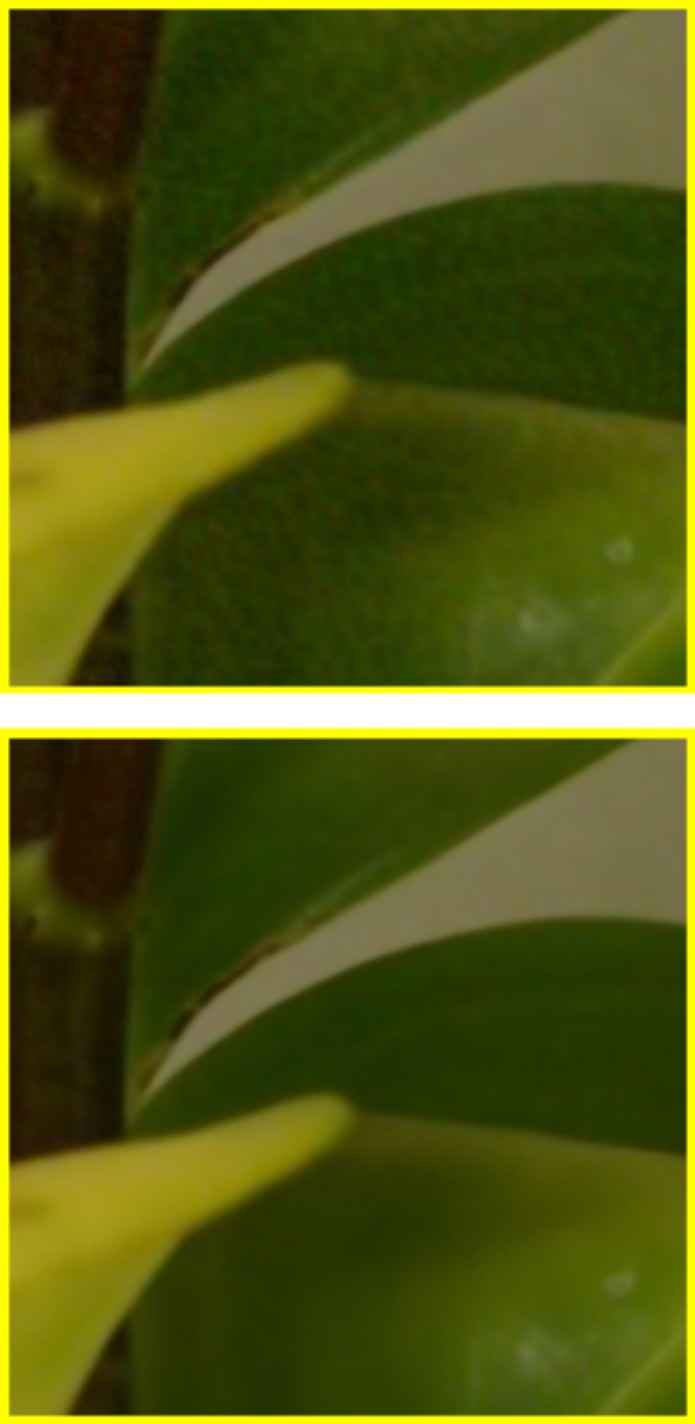}
	}
	\hspace{-6pt}
	\subfloat[]{
		\includegraphics[height=0.48\linewidth]{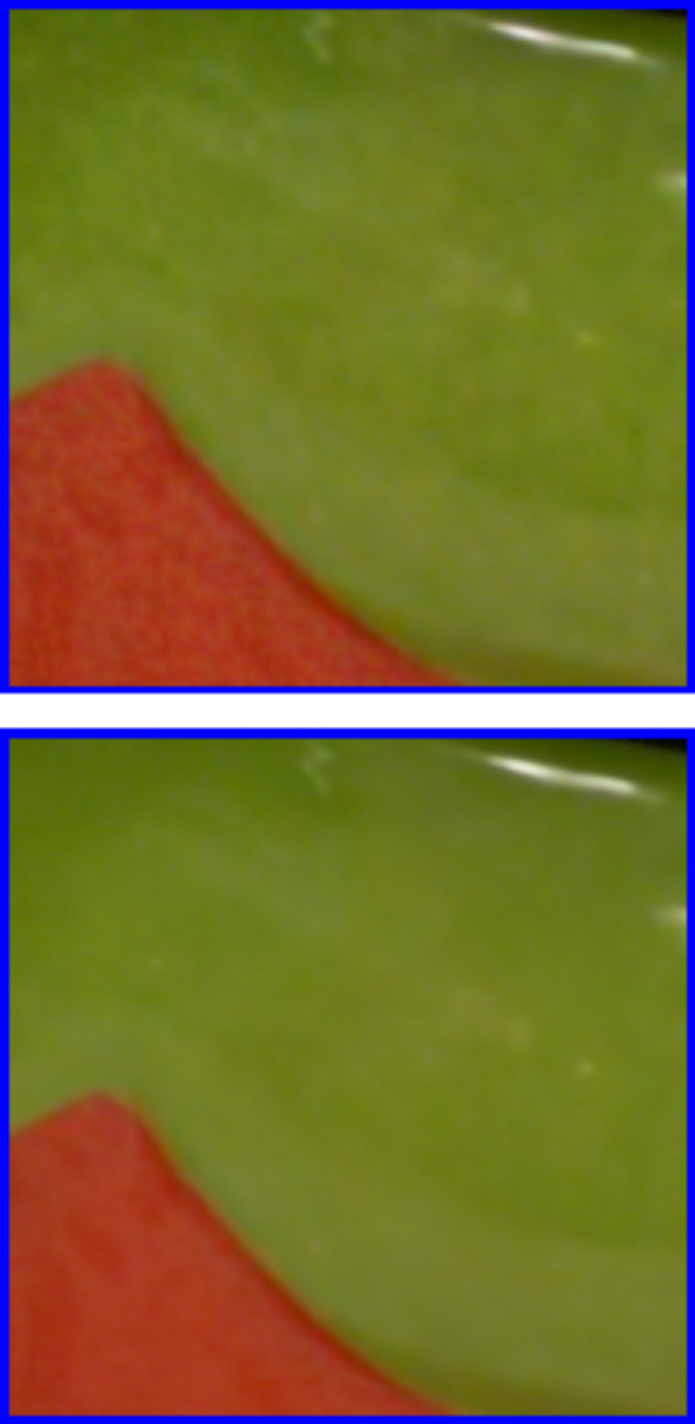}
	}
	\hspace{-6pt}
	\caption{Qualitative comparison between KPNs and AME-KPNs with $5 \!\times\! 5$ kernels on a real-wolrd burst of images within PolyU~\cite{Xu2018PolyU}. Images in the upper and lower panels of (c)-(f) refer to denoised patches by KPNs and AME-KPNs, respectively.}\label{fig:eval_real_data}
\end{figure}

In the next stage, real-world burst images were considered. The noisy reference and the AME-KPN output are respectively shown in Fig.~\ref{fig:eval_real_data}(a) and Fig.~\ref{fig:eval_real_data}(b). For a better view, four patches denoised by KPNs and AME-KPNs are respectively zoomed in the upper and lower panels in Fig.~\ref{fig:eval_real_data}(c)-(f). KPNs produced severe artifacts inside the black triangle and the brown letter ``A'', and on the surface of leaves and flowerpot. Although trained on the synthetic data, owing to the attention mechanism and the residual maps for compensation, our proposed AME-KPNs restored the details of the considered image more efficiently, and in this sense, it possesses a better generalization ability than the original KPNs to real-world burst images.
%
\vspace{-8pt}
\subsection{Ablation Study}\label{sec:ablation_study}
\vspace{-4pt}
In the final stage, an ablation study was performed to verify the individual effectiveness of the channel and spatial attention modules and residual maps; the three novel units within the proposed AME-KPNs, and the results are summarized in Table \ref{tab:ablation_study}. All the six models with $5 \!\times\! 5$ spatially-adaptive kernels were trained on the same synthetic dataset discussed in Section \ref{sec:synthetic} and were evaluated on 100 validation images, generated in the same way as the training data. Particularly, Model 1 refers to the original KPN, and Model 6 is the full version of the proposed AME-KPN, which, as expected, outperformed other models. When either the channel attention module or the spatial attention one was equipped up with KPNs individually, an improvement about 0.6$\sim$0.7 dB in PSNR can be obtained. Moreover, a collaboration of both brought nearly 1 dB increase in PSNR, because of the full use of inter-frame and intra-frame information within the burst images. Besides, the performance improvement introduced by predicting residuals and weight maps simultaneously in order to compensate the coarse output of adaptive convolution becomes clear, when comparing Model 1 and Model 5.
%
\begin{table}[!t]
	\centering
	\caption{Ablation study on the proposed AME-KPNs.}
	\label{tab:ablation_study}
	\resizebox{\linewidth}{!}{
		\begin{tabular}{ccccccc}
			\toprule[1pt]
			Modules           & Model 1		 & Model 2      & Model 3 		& Model 4 	   & Model 5      & Model 6      \\
			\midrule[0.5pt]
			Channel Attention &              & $\checkmark$ &         		& $\checkmark$ &              & $\checkmark$ \\
			Spatial Attention &              &         	    & $\checkmark$  & $\checkmark$ &              & $\checkmark$ \\
			Residual Map  &              &         	    &         		&    	       & $\checkmark$ & $\checkmark$ \\
			\midrule[0.5pt]
			PSNR	          & 37.80        & 38.48 		& 38.39   		& 38.78 	   & 38.63        & 39.67        \\
			\bottomrule[1pt]
		\end{tabular}
	}
\end{table}
\vspace{-10pt}
\section{Conclusion}
\vspace{-8pt}
Novel attention mechanism enhanced kernel prediction networks (AME-KPNs) have been proposed for burst image denoising. The proposed attention modules dynamically aggregate pixel-level information within inter frames and intra frames at multiple scales by refining the feature maps. Moreover, with a large receptive filed, residuals and weight maps are beneficial for compensating the coarse prediction of the kernel based adaptive convolution in an efficient way. The effectiveness of the proposed AME-KPNs in burst image denoising has been verified based on both the synthetic and real-world images, as well as an ablation study.

\bibliographystyle{IEEEbib}
\bibliography{bibliography}
\end{document}